\documentclass[twocolumn,showpacs,preprintnumbers,amsmath,amssymb]{revtex4}

\usepackage{graphicx}
\usepackage{dcolumn}
\usepackage{bm}
\usepackage{xcolor}

\begin{document}

\title{The phase diagram of heavy fermions with Cerium and  Europium ions}

\author{V. Zlati\'c$^{}$} 
\author{I. Aviani$^{}$}%

\affiliation{$^{}$Institute of Physics, Zagreb POB 304, Croatia \\}

\date{\today}

\large
\begin{abstract}
\large

Doniach phase diagram of heavy fermions with Ce and Eu ions is explained by 
the scaling solution of the Anderson model.
At  high temperatures, where the rear earth ions behave as nearly independent local moments (LM) 
the system has a large paramagnetic entropy and its properties are defined  by Kondo temperature, 
$T_K(p)$, where $p$ is the external parameter, like pressure or doping.  
For a given $T_K(p)$, the scaling law allows an estimate of the pressure or doping dependence 
of the coupling constant which is then used to find the dependence of the RKKY temperature 
$T_{RKKY}(p)$  and N\'eel temperature $T_N(p)$ on the control parameter. 
The competition between the on-site Kondo coupling and the off-site RKKY coupling determines 
the mechanism by which the system removes the paramagnetic entropy at low temperatures. 
The pressure-induced change of the ground state is explained by the differences 
in the functional form of $T_K(p)$ and $T_{RKKY}(p)$.  
Our theoretical results capture the main features shown by the Doniach diagram 
of CeRu$_2$Ge$_2$, CeCu$_2$(Ge$_{1-x}$Si$_{x})_2$ or EuCu$_2$(Ge$_{1-x}$Si$_{x})_2$.
\end{abstract}

\maketitle

\begin{figure}[!tb]
\begin{center}
\includegraphics[width=9cm]{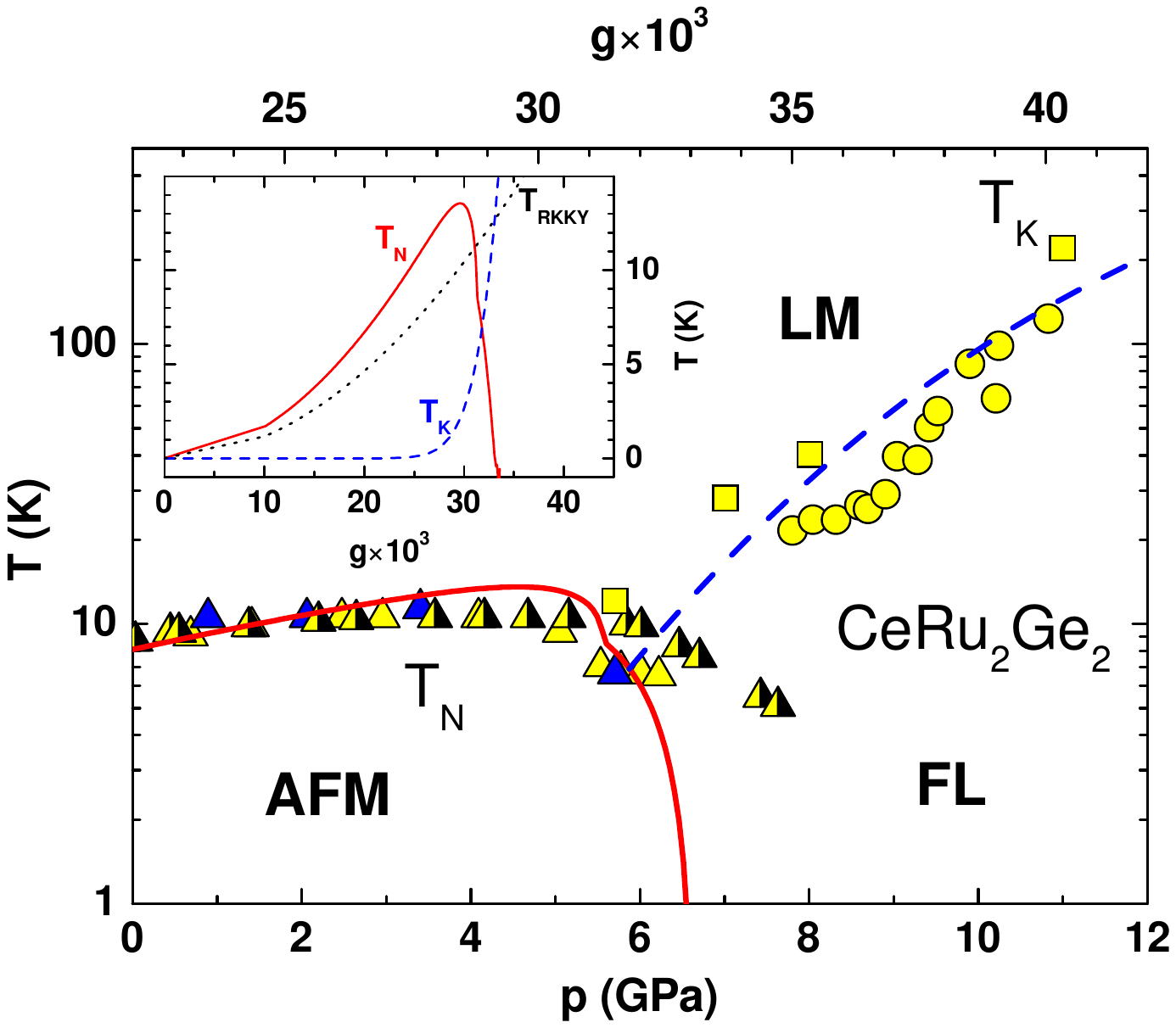} 
\end{center}
\caption{Color on-line. The Kondo scale $T_{K}(p)$ of CeRu$_2$Ge$_2$ 
obtained from data on $\rho(T)$ (squares) and $\alpha(T)$ (circles), 
and the Neel temperature $T_{N}(P)$ obtained from the data 
on $\rho(T)$ (half filled triangles), calorimetric data (open triangles), 
and combined $\rho(T)$ and $\alpha(T)$ data (filled triangles) 
are plotted versus pressure.  
The full and long-dashed lines show $T_{N}(p)$ and $T_{K}(p)$ 
defined by Eqs.~\eqref{eq: Kondo_scaling} and \eqref{eq: T_N}, 
respectively. The inset shows $T_{K}(p)$, $T_{N}(p)$, and $T_{RKKY}(p)$ 
defined by Eq.~\eqref{eq: rkky} (dashed line) plotted versus $g(p)$. 
The model parameters used for the plot and the $p\to g$ mapping  
which defines the upper abscissa, are explained in the text. 
}
                            \label{fig: doniach_Ce} 
\end{figure}

\begin{figure}[!tb]
\begin{center}
\includegraphics[width=10cm]{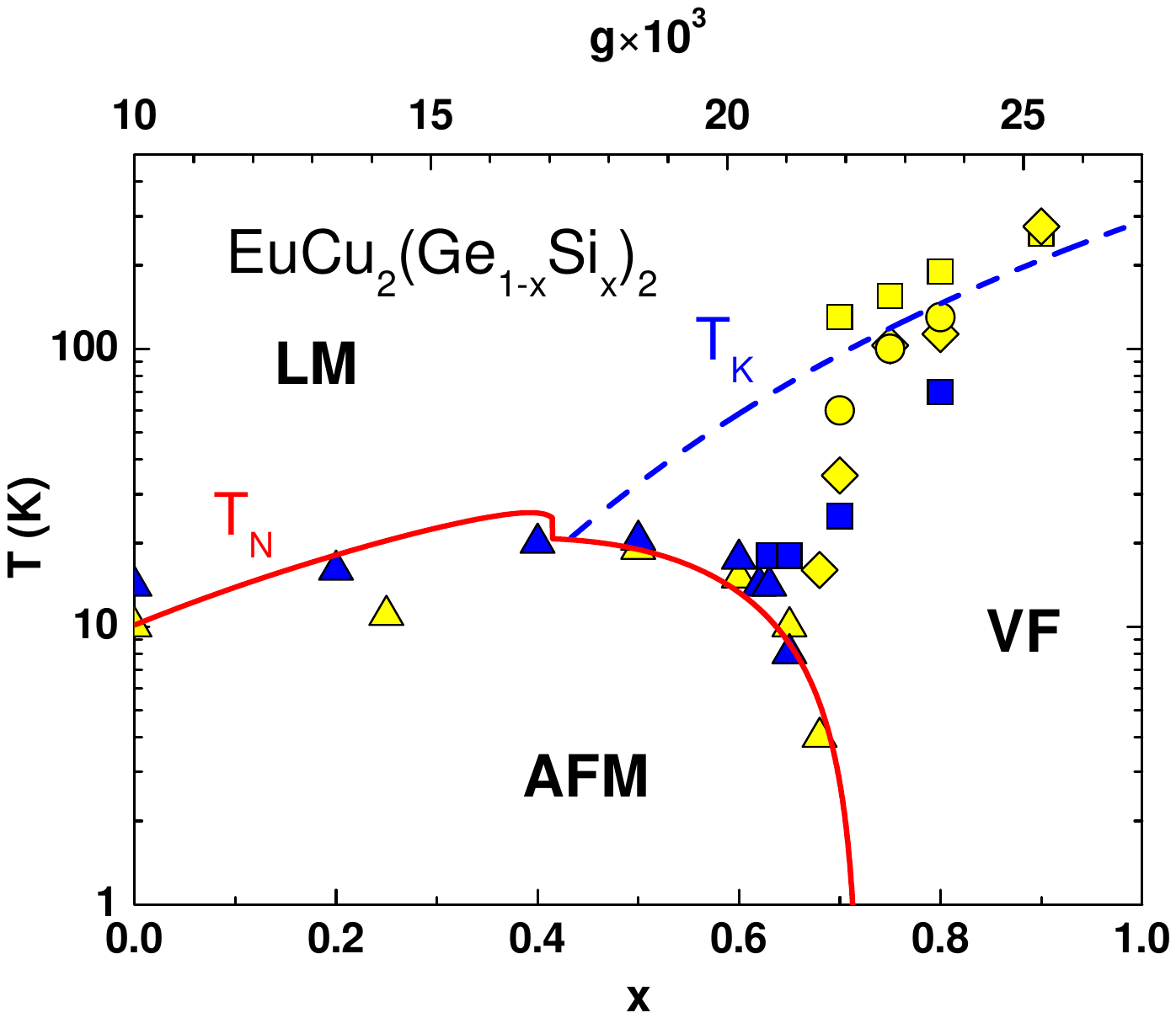} 
\end{center}
\caption{ 
(Color online) The experimental magnetic phase diagram of EuCu$_2$(Ge$_{1-x}$Si$_x$)$_2$. 
is compared with calculated Kondo temperature $T_{K}$ (dashed line) 
and the magnetic ordering temperature $T_{N}$ (full line).The open symbols are the 
data from ref.\cite{fukuda.03} and the bold simbols the data from ref.\cite{hossain.04}. 
The N\'eel temperature (triangles) is obtained from the specific heat anomaly.
The Kondo temperature  is estimated from the thermopower (squares), resistivity (diamonds)
and from the X-ray absorption spectra (circles).
The interaction strength $g=\Gamma /\pi E_{f}$ used to 
calculate  $T_{K}$ and  $T_{N}$ from 
Eqs.~\eqref{eq: Kondo_scaling} and \eqref{eq: T_N}, 
respectively, is shown on upper abscissa.   
}
                            \label{fig: doniach_Eu}
\end{figure}

\paragraph*{\large Introduction}
Heavy fermion (HF) compounds exhibit a number of interesting 
features\cite{doniach.77,hewson.93,wilhelm.02,fukuda.03,wilhelm.04,hossain.04,muramatsu.11} 
due to the competition between the on-site Kondo scattering of conduction electrons on the local 
moments (LM) and the inter-site coupling of the LM induced by the RKKY oscillations.  
The functional form of the response functions of HFs  depends on the relative importance 
of these couplings which are easily changed by a control parameter,  $p$, which could be 
pressure, chemical pressure or magnetic field. 
Plotting the Kondo temperature $T_K(p)$ and N\'eel temperature $T_N(p)$ 
of a given HF compound versus control parameter, yields the Doniach diagram 
which separates the phase space into several characteristic regions\cite{doniach.77,hewson.93}. 
Fig.~\ref{fig: doniach_Ce} shows the Doniach diagram of CeRu$_2$Ge$_2$\cite{wilhelm.04}, 
with pressure as the control parameter,  and Fig.~\ref{fig: doniach_Eu} shows the Doniach diagram 
of EuCu$_2$(Ge$_{1-x}$Si$_{x})_2$\cite{fukuda.03}, with chemical pressure as the control parameter. 
Similar behaviour is also found in CeCu$_2$(Ge$_{1-x}$Si$_{x})_2$\cite{hossain.04}. 
The qualitative features of Doniach diagram have been explained by the Kondo lattice model \cite{doniach.77,kuramoto.09a,savrasov.09} 
but the discussion of specific HF materials and the discussion of the phase diagram at 
the highest pressure is lacking. 

Various phase-space regions of the Doniach diagram, shown in Figs.~\ref{fig: doniach_Ce}  
and ~\ref{fig: doniach_Eu}, exhibit the following characteristic features. 
At high temperatures, the rare earth (RE) ions behave as independent LMs which give rise to Kondo effect, 
so that all the properties of the system at a given pressure are determined by $T_K(p)$. 
The resistivity is, here, a logarithmic function of temperature, the susceptibility is Curie-Weiss like, 
the magnetic moment of 4$f$ electrons is close to what one finds in a free Ce ion,  
and the entropy is dominated by a large paramagnetic contribution, $S_f\simeq k_B \ln N$, 
where $N$ is the effective degeneracy of the LM in a given temperature range\cite{zlatic.05c}. 
Experimentally, $T_K(p)$ is either obtained from transport measurements, 
like thermopower $\alpha(T)$ or electrical resistivity $\rho(T)$, or it is defined by temperature 
at which the entropy drops to half of its high-temperature value\cite{wilhelm.04,fukuda.03,hossain.04}. 
The overall dependence of $T_K(p)$ on the control parmeter is rather smooth, even though the experimental 
values of $T_K(p)$ inferred from different measurements are not exactly the same. 
Theoretically, $T_K$ is defined as the scaling invariant of the Anderson model which we use to describe 
these materials and, for a given coupling constant, it is obtained by solving Eq.~\ref{eq: Kondo_scaling}. 

At low temperatures, a large paramagnetic entropy due to the LM cannot be sustained but 
the mechanism by which the entropy is removed and the nature of the ensuing ground state (GS) 
depend on the relative magnitude of the Kondo and RKKY interactions. 
We distinguish two cases: $ T_K \gg T_{RKKY}$ and  $T_K \ll T_{RKKY}$, 
where  $k_B T_{RKKY}=E_{RKKY}$ is the energy gain due to the antiparallel alignment of the neighbouring LMs 
caused by  the RKKY interaction (in what follows, we set $k_B = 1$).  
The RKKY temperature is related to $T_K$ and $T_N$ but, unlike these temperatures, 
it is not directly discernible in the experimental data. For a given heavy fermion compound, 
the magnitude of $T_{RKKY}(p)$ is estimated a posteriori by model calculations (see Eq.~\ref{eq: rkky}).
At ambient or low pressure we find $T_K(p) < T_{RKKY}(p)$ and the low-entropy state is reached 
at temperature $T_N(p)$ by an AFM transition.  
Above some critical value of the control parameter, $p_c$, we find $T_K(p) >T_{RKKY}(p)$  
and the paramagnetic entropy is not eliminated by the AMF transition but rather by a crossover 
from a LM phase to a heavy fermi liquid (FL). 
The temperature of the crossover is proportional to $T_K(p)$ and the high-pressure behaviour is similar to what 
one finds for an isolated Kondo impurity: at low temperatures, $T\ll T_K$, the conduction electrons screen the LM   
by forming a Kondo singlet and the ensuing GS is a non-degenerate FL\cite{hewson.93}.  

To study the competition between the Kondo interaction and the RKKY interaction,  
and the effect of this competition on the phase diagram of HFs, we use the periodic Anderson model. 
The behaviour of Yb-based intermetallics was explained in Ref.\cite{muramatsu.11} and 
here we discuss the Cerium- and Europium-based intermetallics\cite{fukuda.03,wilhelm.04,hossain.04}.  
The model takes into account the charge transfer between the 4$f$ and $c$-states, 
which is important at high pressure, and also considers the crystal field (CF) splitting, 
which makes  the effective degeneracy of the 4$f$-states pressure and temperature-dependent. 

The paper is organised as follows. First, we introduce the model and the scaling solution. 
At a given pressure, this provides the relationship between the coupling constant $g(p)$ 
and the scaling invariant $T_K(p)$, and reveals the central feature of the Kondo effect, 
namely, the exponential dependence of $T_K(p)$ on $g(p)$. 
The scaling law also yields the renormalised, temperature-dependent coupling constant, 
$g(p,T)$, which is used in the renormalised perturbation theory to study the LM phase. 
By matching the theoretical and experimental values of $T_K(p)$, we obtain the dependence of $g(p)$ 
on the control parameter and calculate $T_{RKKY}(p)$ of a given compound. 
Once we have $T_K(p)$, $g(p)$, and  $g(p,T)$ we can compute the free energy of the LM phase. 
With $T_{RKKY}(p)$ at hand we estimate the free energy of the AFM phase 
and by comparing it with the free energy of the LM phase we find the pressure dependence of $T_N(p)$.  
Finally, the theoretical results are used to discuss the phase diagram of CeRu$_2$Ge$_2$, 
EuCu$_2$(Ge$_{1-x}$Si$_{x})_2$, and similar compounds. 
\\

\paragraph*{\large Model and calculation}
The periodic Anderson model with the CF split 4$f$ states is characterized by 
the unperturbed $c$-band of width $D$, the unrenormalized excitation energy of the 4$f$ states $E_f$,  
the energy gain due to the hybridization of the  4$f$ states with conduction electrons, $\Gamma(p)$, 
and the degeneracies of the crystal field split 4$f$ sates. 
(In the case of independent 4$f$ ions, $\Gamma(p)$ is simply the width of the virtual bound state.)
We consider the model in which the number of $f$-electrons is $n_f$, 
the number of $c$-electrons per site is $n_c$ and assume an infinite $f$-$f$ correlation. 
The degeneracies of the CF states are determined by the point group 
symmetry of the crystal, while the neutron scattering or magnetization data 
provide the splittings. 
For a given $n_c$, $n_f$, and the CF splitting, the low energy 
excitations of the model depend in an essential way\cite{hewson.93} on the dimensionless coupling 
constant $g(p)= \Gamma(p)/\pi|E_f|$. 
In Ce and Eu compounds, $g(p)$ increases with pressure and all the properties change drastically 
due to the exponential dependence of $T_K$ on  $g(p)$, as discussed in Ref.~\onlinecite{zlatic.05c}. 

Since the electrical resistance of heavy fermion compounds in the LM region of the phase space is large, 
we can treat the $4f$ ions as incoherent Kondo scatterers. In that case, the Kondo scale of the model, 
assuming $n_f\simeq 1$, is related to the coupling constant by the 'poor man's scaling'\cite{yamada.84,BCW87}. 
For two excited CF levels which are $N_1$- and $N_2$-fold degenerate, 
and separated from the $N_0$-fold degenerate CF ground state by energies $\Delta_1$ and $\Delta_2$, 
we have the scaling equation\cite{yamada.84,BCW87}, 
\begin{eqnarray}
                            \label{eq: Kondo_scaling}
g(p)&\exp& \left[ -\dfrac{1}{g(p)}\right] =\\
&&
\left(\frac{T_{K}}{D}\right)^{N_0}
\left(\dfrac{T_K+\Delta_1}{D+\Delta_1}\right)^{N_{1}}
\left(\dfrac{T_K+\Delta_2}{D+\Delta_2}\right)^{N_{2}}
, 
\nonumber
~
\end{eqnarray}
where $T_{K}=T_{K}(p)$.  
This  equation holds in the LM regime and to describe CeRu$_2$Ge$_2$ we take 
$N_0=N_1=N_2=2$, $\Delta_1=500$, and $\Delta_2=750$ K \cite{loidl.92}. 

At a given pressure, the properties of the model are calculated 
 by the lowest order perturbation theory with an effective 
temperature-dependent coupling constant $g(p,T)$, 
which is obtained from Eq.~\eqref{eq: Kondo_scaling} by rescaling the $c$-bandwidth 
down to $D\simeq T$.\cite{zlatic.14}
This is equivalent to summing up the most diverging diagrams of the perturbation expansion 
in terms of the bare coupling and yields the correlation functions which are 
universal functions of $T/T_K$\cite{zlatic.14}.
The  results obtained by the renormalized perturbation theory are in a qualitative agreement with 
the NCA\cite{BCW87,zlatic.05c} and the NRG calculations\cite{hewson.93,costi.94,grenzebach.06}, 
which also show that the scaling law holds down to $T<T_K(p)$ 
and it only ceased to be valid for $T\ll T_K(p)$. 

The scaling equation allows us to estimate the pressure dependence of $g(p)$ by 
taking the experimental values of $T_K(p)$ at two different pressures, finding the corresponding 
bare couplings by solving  Eq.~\eqref{eq: Kondo_scaling}, and defining  $g(p)$ at any other 
pressure by a linear interpolation. 
The upper abscissa of the main panel in Fig.~\ref{fig: doniach_Ce} shows $g(p)$ 
obtained for CeRu$_2$Ge$_2$ in such a way, while the inset shows $T_K(p)$ plotted as a function 
of $g(p)$ (long-dashed line). 
The near-exponential dependence of $T_K(p)$ on $g(p)$ is typical of Kondo physics 
and explains the extreme sensitivity of heavy fermions on pressure or chemical pressure. 

In addition to the on-site Kondo coupling, the hybridization between the 4$f$ and $c$-states gives also 
rise to the RKKY spin density oscillations in the $c$-band. This spin density couples to the LMs at 
the neighbouring sites and, if strong enough, it prevents the spin-flip scattering and inhibits the Kondo effect. 
The energy gain due to the RKKY coupling is calculated by the 2nd order perturbation theory in terms 
of the bare coupling\cite{yosida.57}. For Ce ions surrounded by $z$ neighbours at  points {\bf r}, 
this gives 
\begin{eqnarray}
E_{RKKY}(g,{\bf r})
= 18 \pi z S(S+1)| {\cal F}(2{\bf r}\cdot{\bf k}_F)| D \times g^{2}
~,
                     \label{eq: rkky} 
\end{eqnarray} 
where $S$ is the angular momentum of the lowest CF state, 
${\cal F} (\eta ) = [ -\eta\cos\eta + \sin\eta]/\eta^4 $
is the oscillating function, and ${\bf k}_F$ is the Fermi momentum 
of unperturbed $c$-electrons. The value of the coupling constant in Eq.~\eqref{eq: rkky} 
is obtained for a given $T_K(p)$ by solving numerically Eq.~\eqref{eq: Kondo_scaling}. 

The spin density oscillations induced by the RKKY coupling follow, 
like Friedel charge density oscillations, from the Fermi-edge discontinuity of the electron distribution function. 
Thus, they are temperature-dependent  and can be neglected at high temperatures\cite{zlatic.89}. 
At high pressure, where $g(p)$ is large and $T_K(p)$ is huge, the Kondo effect 
reduces the paramagnetic entropy regardless of the RKKY  interaction. 
For $T_{RKKY}\ll T \ll T_K$, the screening of local moments gives rise to the LM-FL crossover. 
On the other hand, at low pressure, the values of $T_K(p)$ decrease exponentially with $g(p)$, 
so that the RKKY coupling, which is a parabolic function of $g(p)$, dominates for $T_K< T < T_{RKKY}$. 
The magnetic field due to the RKKY oscillations inhibits the Kondo effect and, if strong enough, 
it quenches the Kondo scattering and stabilises the  LM on the neighbouring sites before 
the Kondo singlets are formed. In that case, we expect the large entropy of the paramagnetic state 
to be removed at low temperatures by the formation of a magnetically ordered N\'eel state. 
Since Kondo scattering is absent in the magnetically ordered phase, $E_{RKKY}$ 
in Eq.~\eqref{eq: rkky} is calculated with unrenormalised $g(p)$.

The boundary between various characteristic phases of a heavy fermion 
is found by comparing their free energies. Neglecting the entropy of magnetic excitations, 
the free energy of the N\'eel state is 
\begin{eqnarray}
F_N=E_c+E_{f} - E_{RKKY},   \label{eq:F_N}
\end{eqnarray}
where $E_c$ and $E_f$ are the unperturbed internal energies of $c$ and $f$ electrons, respectively, 
and $ E_{RKKY}$ approximates the energy gain due to the alignment of 4$f$ 
moments on the neighbouring sites. 

The free energy in the LM regime is 
\begin{eqnarray}
F_{LM} = E_c + E_f -  E_{fc} - T S_{LM}, \label{eq:F_LM}
\end{eqnarray}
where $E_{fc}$ is the renormalized hybridization energy 
and $S_{LM}$ is the LM entropy. 
The renormalized perturbation theory gives 
$E_{fc} (T ) = \langle H_{cf} \rangle \simeq  g(p,T)T_K$, where 
$H_{cf}$ is the interacting part of the Hamiltonian and $g(p,T )$ 
is obtained from Eq.~\eqref{eq: Kondo_scaling} at $D = T $. 
For $T \gg T_K$, the effective coupling is small, $g(p,T)\ll 1$, 
and the entropy is close to the free-ion value,  $S_{LM} \simeq S_f$.  
At lower temperatures, the effective coupling and $E_{fc}$ grow logarithmically, 
while the entropy decreases.  The renormalized perturbation theory 
yields the approximate relation $S_{LM} \simeq (1-g^3)S_f$.\cite{zlatic.14}

If the paramagnetic entropy of the LM phase is removed by 
the magnetic transition, the N\'eel temperature $T_N$ 
follows from the condition  $F_N=F_{LM}$, such that
$E_{RKKY} = E_{fc} + T_N S_{LM}$. This gives 
\begin{eqnarray}
T_{N}(g) =
\dfrac{E_{RKKY}(g)-g(p,T_N)E_K(g)}{S_{LM}}
~,  
                         \label{eq: T_N}
\end{eqnarray}
where $ S_{LM}$ is the paramagnetic entropy which we approximate by $S_{LM} \simeq S_{f}$. 
A unique determination of $T_N$ for a particular compound requires 
the value of ${\bf k}_F$ in the argument of the oscillating 
function in Eq.~\eqref{eq: rkky}. 
Since this is not known, we adjust the amplitude of ${\cal F} (2 {\bf r\cdot k}_F)$, 
so that the N\'eel temperature at ambient pressure matches the experimental result. 
(Our choice satisfies ${\cal F} (\eta) \geq {\cal F} _{min}$, 
where ${\cal F} _{min} = - 5.06\times 10^{-3}$ 
is the absolute minimum of the oscillating function.) 

Finally, we provide a rough estimate of the phase boundary between the AFM and 
the FL regions of the phase space, where  $T_N\ll  T_K$. 
We approximate the free energy of the FL phase as 
\begin{eqnarray}
F_0=E_c+E_f - E_K - TS_0, \label{eq:F_0}
\end{eqnarray} 
where $S_0\simeq T\gamma \simeq (\pi^2 /3V_0)(T/T_K)$ 
is the entropy of heavy fermions in the FL regime and 
$V_0$ is the unit cell volume\cite{zlatic.08a}. 
The condition $F_0=F_N$ gives $E_{RKKY} = E_K+T_N S_0$, such that 
\begin{eqnarray}
T_{N}^2 \simeq T_K (T_{RKKY} - T_{K}) ~.  
                         \label{eq: T_N0}
\end{eqnarray}
Assumong at critical pressure $T_N(p_c)=0$,  
gives the approximate result, $T_N(p) \simeq \sqrt{p - p_c}$.  
\\

\paragraph*{\large Discussion}
Before presenting the results which show how the competition between Kondo and 
RKKY interactions gives rise to Doniach diagram, we discuss briefly the characteristic 
features of the Anderson model in various parts of the phase space, 
using the parameters relevant for CeRu$_2$Ge$_2$ and similar intermetallics.\cite{wilhelm.04,zlatic.05c}.

At low (ambient) pressure and above 400 K, the low-laying CF states of Cerium ions 
are occupied with equal probability, so that the conduction electrons scatter on the six-fold degenerate LMs. 
This gives rise to the resistivity and thermopower which are logarithmic functions of temperature 
(with a negative slope).\cite{wilhelm.04,zlatic.05c}
Below 400 K the excited CF states depopulate and around $T_{\Delta}\simeq 350$ K
 there is a crossover to a new LM regime, where the 4$f$ state behaves as an effective CF doublet.   
This LM-LM crossover is indicated in the resistivity and thermopower data by the high-temperature 
maxima (see Figs. 2 and 3 in Ref.\cite{wilhelm.04}).
The thermopower maximum is particularly pronounced, as $\alpha (T)$ 
drops around $T_{\Delta}$ from positive to negative values.  
For $T_K <T < T_{\Delta}$, the Kondo scattering on effective CF doublets gives rise to 
the resistivity and thermopower which increase towards their low-temperature maxima. 
The associated Kondo scale, inferred from the low-temperature maximum of $\alpha(T)$ 
or $\rho(T)$, is very small\cite{zlatic.05c}, such that $T_K\ll T_{RKKY}$. 
Thus, it is not surprising that at ambient or low pressure, the paramagnetic entropy of CeRu$_2$Ge$_2$ 
is removed at low temperatures by an AFM transition, as indicated by a large specific heat anomaly 
and the discontinuity in the slope of $\rho (T )$ and $\alpha (T)$.  
The magnetic moment of Ce ions in the ordered state is much smaller 
than of a free Ce ion but our analysis shows that this reduction is a CF effect and 
it is not due to the Kondo screening. 

The model calculations and the experimental data show that $T_N(p)$ increases gradually with pressure up to 
a maximum and than drops rapidly. For $T_N \simeq T_K$, a precise estimate of $T_N$ from transport data 
becomes difficult, because the two maxima which characterise $\alpha (T)$ and $\rho (T)$ are
merged at higher pressure into a single broad maximum, such that a weak discontinuity 
of the slope is difficult to measure.
Above the critical pressure, $p_c\simeq $ 6 GPa, we find $T_N < T_K$ and 
Kondo effect inhibits the formation of a magnetically ordered state. 

At large pressure,  $p\geq p_c$,  we find $T_{RKKY}\ll T_K $, so that 
the RKKY interaction can be neglected throughout the LM regime. 
As temperature is reduced, the Kondo scattering leads gradually to the screening of the LM 
and, for $T\ll T_K$, the coherent state forms out of Kondo singlets\cite{burdin.09}. 
The CeRu$_2$Ge$_2$ and similar compounds behave at low temperatures and $p\geq p_c$ as a heavy FL 
with an enhanced Pauli-like susceptibility $\chi$, and a large specific heat coefficient, $\gamma=C_V/T$. 
The calculations for the periodic Anderson model show\cite{zlatic.08,burdin.09} that the enhancement 
of $\chi$ and $\gamma$ scales with $T_K$, i.e., 
Kondo temperature provides the relevant energy scale at high and low temperatures. 
In the FL region of the phase space, the transport coefficients are given by simple powers of $T/T_K$ 
and the system is characterised by various universal ratios, like the Wilson ratio $\chi/\gamma$, 
the Kadowaki-Woods ratio $\rho / \gamma$ or the $q$-ratio $\alpha / \gamma$. 

The comparison between the experimental  and theoretical Doniach diagrams of 
CeRu$_2$Ge$_2$ is provided by Fig.~\ref{fig: doniach_Ce}, where the dashed line 
shows $T_K(p)$ calculated by the scaling theory and the full line is $T_N$. 
The values of $g(p)$ used in the calculations are given on the upper abscissa and 
$T_{RKKY}$ is evaluated for the lowest CF doublet (S = 1/2), 
with D = 4 eV, z = 6, and $|F(\eta)| = -7.86 \times 10^{-4}$. 

The gradual increase of $T_N(p)$ above the ambient pressure is due to 
the fact that $T_K(p)$  is exponentially smaller than $T_{RKKY}(p)$, so it can be 
neglected in Eq.~\ref{eq: T_N} (see inset in Fig.~\ref{fig: doniach_Ce}). 
Above certain pressure,  the exponential growth of $T_K(p)$ brings $T_N(p)$ 
to a maximum and, then, reduces it sharply to zero. 
The asymmetric shape of $T_N(p)$ is due to different functional 
forms of $T_K(p)$ and  $T_{RKKY}(p)$. 

The scaling solution of an eight-fold degenerate Anderson model  
explains the phase diagram of EuCu$_2$(Ge$_{1-x}$Si$_x$)$_2$,  
where Silicon doping provides the chemical pressure which drives this system 
from an antiferromagnet to a valence fluctuator.\cite{fukuda.03,hossain.04}
The comparison with the experiment is shown in Fig.~\ref{fig: doniach_Eu}, 
where $T_N$ (full line) and $T_K$ (dashed line) are plotted versus Si concentration 
(lower abscissa). The corresponding values of $g(x)$ are given on the upper abscissa 
and the calculations are carried out following the same steps as in the case 
of CeRu$_2$Ge$_2$, taking S=7/2, D = 4 eV, z = 6, and $|F(\eta)| = -8.51 \times 10^{-4}$. 

The calculated phase boundary in Fig.~\ref{fig: doniach_Eu} exhibits 
the same generic features as the experimental one. 
The difficulty is that our calculations take into account the spin-flip scattering of $c$-electrons 
on the eight-fold degenerate Eu$^{2+}$ (4$f^7$) ions but neglect the fluctuations  
between Eu$^{2+}$ and Eu$^{3+}$ (4$f^6$) configurations. 
These fluctuations become important for large $x$ but to include them one would have to go 
beyond the scaling theory and consider a modified Hamiltonian which includes the Falicov-Kimball term\cite{freericks.03}. 
Furthermore, the characteristic temperatures obtained from various experiments on 
EuCu$_2$(Ge$_{1-x}$Si$_x$)$_2$ differ by a factor of 2 or 3, 
and the values of $T_K$ inferred from the experiment have a large error bar. \\

\paragraph*{\large Summary}
We described the phase diagram of heavy fermions with Ce and Eu ions 
using the scaling solution of the periodic Anderson model. 
At  high temperatures, where the system is in the LM phase with large paramagnetic entropy, 
it is completely characterised by its Kondo temperature.  
The scaling law allows us to estimate the pressure or doping dependence of the coupling constant 
which we then use to find the pressure dependence of $T_{RKKY}$ and $T_N$. 
The competition between the on-site Kondo coupling and the off-site RKKY coupling determines 
the mechanism by which the compound removes the paramagnetic entropy at low temperatures. 
The pressure-induced change in the ground state is explained by the differences 
in the  functional form of $T_K(p)$ and $T_{RKKY}(p)$.  
Our theoretical results capture the main experimental features shown by the phase diagram of 
CeRu$_2$Ge$_2$, EuCu$_2$(Ge$_{1-x}$Si$_x$)$_2$ and 
other heavy fermions in which pressure or chemical pressure increase the coupling constant.

\paragraph*{Acknowledgment} 
We acknowledge support by the Ministry of Science of Croatia under the bilateral agreement with the USA 
on scientific and technological cooperation, Project No. 2/2019.

\newpage

NOTE:\\
Dear Editor,

The plots of Neel and Kondo temperatures versus pressure or chemical pressure 
are often found in the experimental literature on heavy fermions (HF), where they 
are referred to as Doniach diagrams.   
However, the actual calculations of Doniach diagram of any specific material have 
never been presented.  Our paper calculates the Doniach diagram of HFs with Ce 
and Eu ions and explains the pressure dependence of Kondo and Neel temperature 
of these systems. 
\\
We obtain Doniach diagram by using the scaling solution of the Anderson model in 
which a given HF is characterised by its Kondo scale $T_K(P)$. Here, P is the control 
parameter, like pressure or chemical pressure.  The scaling law allows us to find the 
dimensionless coupling $g(P)$ as a function of pressure, so we can calculate the 
pressure dependence of the RKKY temperature $T_{RKKY}(P)$. 
The Neel temperature $T_N$ is obtained by equating the free energies of the paramagnetic 
and the antiferromagnetic phases. Our  solution captures the main experimental features 
shown by CeRu$_2$Ge$_2$ and other HFS in which pressure increases the coupling constant 
and makes them less magnetic. Our approach also explains the behaviour of Eu-based compounds, 
which become non-magnetic at high pressure. 
 \\
We believe, the paper meets all the criteria required for the publication in the PRB.

Sincerely yours,
V.Z. and I.A.

\end{document}